\documentclass[a4paper]{jpconf}

\begin{document}

\title{Multiloop calculations in supersymmetric theories with the higher
covariant derivative regularization}

\author{K V Stepanyantz}

\address{Department of Theoretical Physics, Physical Faculty,
Moscow State University, Moscow, Russia\\
E-mail: stepanyantz@mail.ru}

\begin{abstract}
Most calculations of quantum corrections in supersymmetric
theories are made with the dimensional reduction, which is a
modification of the dimensional regularization. However, it is
well known that the dimensional reduction is not self-consistent.
A consistent regularization, which does not break the
supersymmetry, is the higher covariant derivative regularization.
However, the integrals obtained with this regularization can not
be usually calculated analytically. We discuss application of this
regularization to the calculations in supersymmetric theories. In
particular, it is demonstrated that integrals defining the
$\beta$-function are possibly integrals of total derivatives. This
feature allows to explain the origin of the exact NSVZ
$\beta$-function, relating the $\beta$-function with the anomalous
dimensions of the matter superfields. However, integrals for the
anomalous dimension should be calculated numerically.
\end{abstract}


\section{Introduction}
\hspace{\parindent}

In order to deal with divergent expressions in the quantum field
theory, it is necessary to regularize a theory. A proper choice of
a regularization can simplify the calculations or reveal some
features of quantum corrections. Most calculations in the quantum
field theory were made with the dimensional regularization
\cite{tHooft} in $\overline{MS}$-scheme \cite{MS}. However, the
dimensional regularization breaks the supersymmetry and is not
convenient for calculations in supersymmetric theories. That is
why most calculations in supersymmetric theories were made with
the dimensional reduction \cite{Siegel}. For example, the
$\beta$-function in supersymmetric theories was calculated up to
the four-loop approximation \cite{1Loop,2Loop,3Loop,3Loop2,4Loop}.
After a special redefinition of the coupling constant
\cite{JackJonesNorth,JackJonesPickering} the result coincides with
the exact NSVZ $\beta$-function, proposed in Refs.
\cite{NSVZ1,NSVZ2,NSVZ3,SQED_Beta_Derivation}.

However, it is well known that the dimensional reduction is not
self-consistent \cite{Siegel2}. Removing the inconsistencies one
breaks the explicit supersymmetry \cite{Chochia,Stockinger}. Then
the supersymmetry can be broken by quantum corrections in higher
loops \cite{Avdeev,AvdeevVladimirov,Velizhanin}. In the $N=2$ SYM
theory this already occurs in the three-loop approximation
\cite{Avdeev,Velizhanin}, while in the $N=4$ SYM theory the
supersymmetry is not broken even in the four-loop approximation
\cite{VelizhaninN4}. Thus, a problem of regularization in
supersymmetric theories is rather nontrivial \cite{Jones_Reg}.

For supersymmetric theories one can use the higher covariant
derivative regularization, proposed by A.A.Slavnov
\cite{Slavnov1,Slavnov2}. Different versions of this
regularization for supersymmetric theories, which do not break the
supersymmetry, were proposed in \cite{Krivoshchekov,West_Paper}.
Unlike the dimensional reduction, the higher covariant derivative
regularization is consistent. However, it was not often applied to
explicit calculations of quantum corrections, because the
corresponding integrals have very complicated structure, and it is
not easy to calculate them analytically, especially in higher
loops. Moreover, some theoretical subtleties can raise nontrivial
questions even in the simplest calculations
\cite{Ruiz,Asorey,Bakeyev}.

However, we argue that for supersymmetric theories this
regularization has some very attractive features and can be used
for the calculations. Namely, the integrals defining the
$\beta$-function are integrals of double total derivatives
\cite{3LoopHEP,Smilga,SD,PhysLettSUSY}, and one of them can be
calculated analytically. As a result, the $\beta$-function is
related with the anomalous dimension, producing the exact NSVZ
$\beta$-function without redefinition of the coupling constant. In
this paper we demonstrate how this can proved in $N=1$ SQED in all
loops and for the general renormalizable $N=1$ SYM in the two-loop
approximation.

\section{Quantum corrections in N=1 SQED, regularized by higher derivatives}

\subsection{Higher derivative regularization}
\hspace{\parindent}

The action of the massless $N=1$ SQED in terms of superfields
\cite{West,Buchbinder} is written as

\begin{equation}\label{SQED_Action}
S = \frac{1}{4 e^2} \mbox{Re}\int d^4x\,d^2\theta\,W_a C^{ab} W_b
+ \frac{1}{4}\int d^4x\, d^4\theta\, \Big(\phi^* e^{2V}\phi
+\widetilde\phi^* e^{-2V}\widetilde\phi\Big).
\end{equation}

\noindent The theory is regularized by adding the term with the
higher derivatives

\begin{equation}\label{Regularized_SQED_Action}
S_\Lambda = \frac{1}{4 e^2} \mbox{Re}\int d^4x\,d^2\theta\,W_a
C^{ab} \Big(R\Big(\frac{\partial^{2}}{\Lambda^{2}}\Big)-1\Big)
W_b,
\end{equation}

\noindent where $R(0) = 1$ and $R(\infty) = \infty$. For example,
it is possible to choose $R = 1 + \partial^{2n}/\Lambda^{2n}$. The
gauge is fixed by adding

\begin{equation}
S_{\mbox{\scriptsize gf}} = - \frac{1}{64 e^2}\,\int
d^4x\,d^4\theta\, \Big(V R\Big(\frac{\partial^2}{\Lambda^2}\Big)
D^2 \bar D^2 V + V R\Big(\frac{\partial^2}{\Lambda^2}\Big) \bar
D^2 D^2 V\Big).
\end{equation}

\noindent Then the propagator will contain large degrees of the
momentum in the denominator, and all loop diagrams beyond the
one-loop approximation become convergent. The remaining one-loop
diagrams are regularized by inserting the Pauli--Villars
determinants \cite{Slavnov_Book}

\begin{equation}
\prod\limits_{I}\Big(\int D\phi_{I}^* D\phi_{I}
e^{iS_{I}}\Big)^{-c_I}
\end{equation}

\noindent into the generating functional, where

\begin{equation}
S_I = \frac{1}{4} \int d^4x\,d^4\theta\,\Big(\phi_I^* e^{2V}
\phi_I + \widetilde\phi_I^* e^{-2V} \widetilde\phi_I\Big) +
\Big(\frac{1}{2}\int d^4x\,d^4\theta\,M_I \phi_I\widetilde\phi_I
+\mbox{h.c.}\Big)
\end{equation}

\noindent and $\sum c_I = 1$, $\sum c_I M_I^2 = 0$. It is
important that the masses $M_I$ are proportional to the parameter
$\Lambda$.

\subsection{Three-loop $\beta$-function}
\hspace{\parindent}

In order to find the $\beta$-function we consider

\begin{equation}\label{D_Definition}
\Gamma^{(2)}_V = - \frac{1}{16\pi} \int
\frac{d^4p}{(2\pi)^4}\,d^4\theta\, V(-p)\,\partial^2\Pi_{1/2}
V(p)\, d^{-1}(\alpha,\mu/p)
\end{equation}

\noindent and calculate

\begin{equation}\label{We_Calculate}
\frac{d}{d\ln \Lambda}\,
\Big(d^{-1}(\alpha_0,\Lambda/p)-\alpha_0^{-1}\Big)\Big|_{p=0} = -
\frac{d\alpha_0^{-1}}{d\ln\Lambda} =
\frac{\beta(\alpha_0)}{\alpha_0^2}.
\end{equation}

\noindent In the three-loop approximation the result can be
written as ($R_k \equiv R(k^2/\Lambda^2)$)

\begin{eqnarray}\label{SQED_Result}
&& \frac{\beta(\alpha_0)}{\alpha_0^2} = 2\pi \frac{d}{d\ln\Lambda}
\sum\limits_{I} c_I \int \frac{d^4q}{(2\pi)^4}
\frac{\partial}{\partial q^\mu} \frac{\partial}{\partial q_\mu}
\frac{\ln(q^2+M^2)}{q^2} + 4\pi \frac{d}{d\ln\Lambda} \int
\frac{d^4q}{(2\pi)^4} \frac{d^4k}{(2\pi)^4} \frac{e^2}{k^2 R_k^2}
\nonumber\\
&& \times \frac{\partial}{\partial q^\mu} \frac{\partial}{\partial
q_\mu} \Bigg(\frac{1}{q^2 (k+q)^2} - \sum\limits_I c_I
\frac{1}{(q^2+M_I^2)((k+q)^2 + M_I^2)}\Bigg) \Bigg[ R_k
\Big(1+\frac{e^2}{4\pi^2}\ln\frac{\Lambda}{\mu}\Big) \nonumber\\
&& - 2 e^2 \Bigg(\int \frac{d^4t}{(2\pi)^4}\,\frac{1}{t^2 (k+t)^2}
-\sum\limits_J c_J \int \frac{d^4t}{(2\pi)^4} \frac{1}{(t^2+M_J^2)
((k+t)^2+M_J^2)} \Bigg)\Bigg] \nonumber\\
&& + 4\pi \frac{d}{d\ln\Lambda} \int \frac{d^4q}{(2\pi)^4}
\frac{d^4k}{(2\pi)^4} \frac{d^4l}{(2\pi)^4} \frac{e^4}{k^2 R_k l^2
R_l} \frac{\partial}{\partial q^\mu} \frac{\partial}{\partial
q_\mu}\Bigg\{\Bigg( - \frac{2 k^2}{q^2 (q+k)^2 (q+l)^2 (q+k+l)^2}
\nonumber\\
&& + \frac{2}{q^2(q+k)^2(q+l)^2}\Bigg) - \sum\limits_I c_I \Bigg(-
\frac{2(k^2 + M_I^2)}{(q^2+M_I^2) ((q+k)^2+M_I^2) ((q+l)^2+M_I^2)}
\nonumber\\
&& \frac{1}{((q+k+l)^2+M_I^2)} + \frac{2}{(q^2+M_I^2)
((q+k)^2+M_I^2)((q+l)^2+M_I^2)} - \frac{1}{(q^2+M_I^2)^2}
\nonumber\\
&& \times \frac{4M_I^2}{((q+k)^2+M_I^2) ((q+l)^2+M_I^2)}
\Bigg)\Bigg\}.\vphantom{\frac{1}{2}}
\end{eqnarray}

\noindent From this expression we see that the $\beta$-function is
defined by the integrals of (double) total derivatives
\cite{3LoopHEP,Smilga,SD}. Such a structure allows to calculate
one of the loop integrals analytically and, thus, to reduce a
number of integrations. This can be done using the identity

\begin{equation}\label{Integral}
\int \frac{d^4q}{(2\pi)^4} \frac{\partial}{\partial q^\mu}
\frac{\partial}{\partial q_\mu} \Big(\frac{f(q^2)}{q^2}\Big) =
\lim\limits_{\varepsilon\to 0}\int\limits_{S_\varepsilon}
\frac{dS_\mu}{(2\pi)^4} \frac{(-2) q^\mu f(q^2)}{q^4} =
\frac{1}{4\pi^2} f(0),
\end{equation}

\noindent where $f$ is a nonsingular function which rapidly tends
to 0 at the infinity. It is equivalent to the identity

\begin{equation}
\int \frac{d^4q}{(2\pi)^4} \frac{1}{q^2} \frac{d}{dq^2} f(q^2) =
\frac{1}{16\pi^2} \Big(f(\infty) - f(0)\Big)= -\frac{1}{16\pi^2}
f(0).
\end{equation}

\noindent (This is a total derivative in the four-dimensional
spherical coordinates.) Taking the integrals in Eq.
(\ref{SQED_Result}) and comparing the result with the expression
for the two-loop anomalous dimension of the matter superfield

\begin{eqnarray}
&&\hspace*{-5mm} \gamma(\alpha_0) = - 2e^2 \int
\frac{d^4k}{(2\pi)^4} \frac{d}{d\ln\Lambda} \frac{1}{k^4 R_k^2}
\Bigg[R_k \Big(1+\frac{e^2}{4\pi^2}\ln\frac{\Lambda}{\mu} \Big) -
\int
\frac{d^4t}{(2\pi)^4}\,\frac{2 e^2}{t^2 (k+t)^2}\\
&&\hspace*{-5mm} + \sum\limits_I c_I \int \frac{d^4t}{(2\pi)^4}\,
\frac{2 e^2}{(t^2+M_I^2) ((k+t)^2+M_I^2)} \Bigg] - \int
\frac{d^4k}{(2\pi)^4} \frac{d^4l}{(2\pi)^4}\,\frac{d}{d\ln
\Lambda}\frac{4 e^4 k_\mu l_\mu}{k^4 R_k\,l^4 R_l
(k+l)^2},\nonumber
\end{eqnarray}

\noindent in the considered approximation  we obtain the exact
NSVZ $\beta$-function for the $N=1$ SQED
\cite{SQED_Beta_Derivation}:

\begin{equation}
\beta(\alpha_0) =
\frac{\alpha_0^2}{\pi}\Big(1-\gamma(\alpha_0)\Big) +
O(\alpha_0^5).
\end{equation}

\noindent This equality of the well-defined integrals is obtained
without any redefinitions of the coupling constant. Thus, with the
higher derivative regularization the NSVZ scheme can be naturally
defined. However, it is much more complicated problem to calculate
analytically the integrals for the anomalous dimension. Possibly,
in the lowest orders this can be done analytically, but it seems
that in higher loops for this purpose one should use numerical
methods.

\subsection{The exact result}
\hspace{\parindent}

For the $N=1$ SQED it is possible to demonstrate that the features
discussed in the previous section take place in all loops. In
particular, the integrals defining the $\beta$-function are
integrals of double total derivatives, and the $\beta$-function
coincides with the NSVZ expression without a redefinition of the
coupling constant \cite{NuclPhys}.

For this purpose, first, we make the substitution

\begin{equation}\label{Substitution}
V \to \bar\theta^a\bar\theta_a \theta^b\theta_b\equiv \theta^4,
\end{equation}

\noindent which allows to extract the function $d^{-1}$ in Eq.
(\ref{D_Definition}). Then we will try to present the sum of
Feynman diagrams as integrals of total derivatives, which in the
coordinate representation are given by

\begin{equation}
\mbox{Tr} \Big([x^\mu, \mbox{Something}]\Big) = 0.
\end{equation}

We start with the expression for the part of the effective action
corresponding to the two-point Green function of the gauge
superfield \cite{NuclPhys}

\begin{equation}\label{Effective_Action}
\Delta\Gamma^{(2)}_{\bf V} = \Big\langle - 2i \Big(\mbox{Tr}({\bf
V} J_0 *)\Big)^2 - 2i \mbox{Tr}({\bf V} J_0
* {\bf V} J_0 *) - 2i \mbox{Tr}({\bf V}^2 J_0 *)\Big\rangle
+ \mbox{terms with $\widetilde *$} + (PV),
\end{equation}

\noindent where

\begin{equation} * \equiv
\frac{1}{1-(e^{2V}-1)\bar D^2 D^2/16\partial^2}, \qquad \widetilde
* = \frac{1}{1-(e^{-2V}-1)\bar D^2 D^2/16\partial^2}
\end{equation}

\noindent encode sequences of vertexes and propagators on the
matter line, and $J_0 = e^{2V} \bar D^2 D^2/16\partial^2$ is the
effective vertex. $(PV)$ denotes contributions of the
Pauli--Villars fields. The first term in Eq.
(\ref{Effective_Action}) is a sum of diagrams in which external
lines are attached to different loops of the matter superfields.
The second term is a sum of diagrams in which external lines are
attached to a single line of the matter superfields. The last term
is not transversal. The sum of such terms vanishes due to the Ward
identities.

After substitution (\ref{Substitution}) and some algebraic
transformations \cite{NuclPhys} the first term in Eq.
(\ref{Effective_Action}) gives the contribution

\begin{equation}\label{First_Term}
-2i \frac{d}{d\ln\Lambda} \Big\langle \Big(\mbox{Tr} \Big(
-2\theta^c\theta_c \bar\theta^d [\bar\theta_d, \ln(*) -
\ln(\widetilde *)] + i \bar\theta^c (\gamma^\nu)_c{}^d \theta_d
[y_\nu^*, \ln(*) - \ln(\widetilde *)]\Big) +
(PV)\Big)^2\Big\rangle.
\end{equation}

\noindent Similarly, the second term in Eq.
(\ref{Effective_Action}) gives

\begin{equation}\label{Second_Term}
i\frac{d}{d\ln\Lambda} \mbox{Tr}\Big\langle\theta^4 \Big[y_\mu^*,
\Big[(y^\mu)^*,\ln(*) +\ln(\widetilde *) \Big] \Big]\Big\rangle
+(PV) - \mbox{terms with a $\delta$-function}.
\end{equation}

\noindent The third term in Eq. (\ref{Effective_Action}) vanishes
after substitution (\ref{Substitution}). From expressions
(\ref{First_Term}) and (\ref{Second_Term}) we see that in all
orders the $\beta$-function is given by integrals of double total
derivatives. A different method to see this \cite{Smilga} is based
on the covariant Feynman rules in the background field method
\cite{Grisaru1,Grisaru2}.

Expressions (\ref{First_Term}) and (\ref{Second_Term}) can be
calculated explicitly in the three-loop approximation. The result
coincides with Eq. (\ref{SQED_Result}).

Terms with the $\delta$-function in Eq. (\ref{Second_Term}) appear
due to the identity

\begin{equation}
[x^\mu,\frac{\partial_\mu}{\partial^4}] =
[-i\frac{\partial}{\partial p_\mu}, -\frac{ip^\mu}{p^4}] = -2\pi^2
\delta^4(p_E) = -2\pi^2 i \delta^4(p).
\end{equation}

\noindent Due to this $\delta$-function one of loop integrals can
be calculated and a number of integrations is reduced.
Qualitatively, we consider all diagrams in which two external
gauge lines are attached to the same graph. Integration of the
$\delta$-function corresponds to cutting a matter line in this
graph \cite{Smilga}. This gives diagrams with two external matter
lines, defining the anomalous dimension. Thus, the
$\beta$-function in a certain loop is reduced to the anomalous
dimension in the previous loop. These arguments can formulated
rigorously \cite{NuclPhys} in all orders and allow to obtain the
exact NSVZ $\beta$-function

\begin{equation}\label{NSVZ_SQED}
\beta(\alpha) = \frac{\alpha^2}{\pi}\Big(1-\gamma(\alpha)\Big).
\end{equation}

\noindent Note that deriving Eq. (\ref{NSVZ_SQED}) one does not
redefine the coupling constant, as in the case of the dimensional
reduction \cite{JackJonesNorth,JackJonesPickering}. Therefore,
with the higher derivative regularization we can naturally define
the NSVZ scheme for the $N=1$ SQED.

\section{Two-loop $\beta$-function with the higher covariant
derivative regularization in the non-Abelian case}
\hspace{\parindent}

Let us consider a general renormalizable $N=1$ supersymmetric
Yang--Mills theory with a gauge group $G$ and matter superfields
$\phi_i$ in a representation $R$, in the massless limit:

\begin{eqnarray}\label{SYM_Action}
&& S = \frac{1}{2 e^2} \mbox{Re}\,\mbox{tr}\int
d^4x\,d^2\theta\,W_a C^{ab} W_b + \frac{1}{4}\int d^4x\,
d^4\theta\, (\phi^*)^i (e^{2V})_i{}^j\phi_j +\nonumber\\
&&\qquad\qquad\qquad\qquad\qquad\qquad\qquad\qquad +
\Bigg(\frac{1}{6} \int d^4x\, d^2\theta\,\lambda^{ijk} \phi_i
\phi_j \phi_k + \mbox{h.c.}\Bigg)\qquad
\end{eqnarray}

\noindent This theory is invariant under the gauge transformation
if

\begin{equation}\label{Lambda_Invariance}
(T^A)_m{}^{i}\lambda^{mjk} + (T^A)_m{}^{j}\lambda^{imk} +
(T^A)_m{}^{k}\lambda^{ijm} = 0.
\end{equation}

\noindent Below we assume that this condition is satisfied. In
order to introduce the higher covariant derivative regularization
and calculate the $\beta$-function in this case we will use the
background field method. In the supersymmetric case
\cite{West,Buchbinder} we split the gauge superfield (which is
below denoted by $V'$) into the quantum part $V$ and the
background field $\mbox{\boldmath${\Omega}$}$ according to the
prescription $e^{2V'} \equiv e^{\mbox{\boldmath${\scriptstyle
\Omega}$}^+} e^{2V} e^{\mbox{\boldmath${\scriptstyle \Omega}$}}$.
Then the gauge can be fixed without breaking the background gauge
invariance:

\begin{equation}\label{Gauge_Fixing}
S_{\mbox{\scriptsize gf}} = - \frac{1}{32 e^2}\,\mbox{tr}\,\int
d^4x\,d^4\theta\, \Big(V \mbox{\boldmath$D$}^2
\bar{\mbox{\boldmath$D$}}^2  V + V \bar {\mbox{\boldmath$D$}}^2
\mbox{\boldmath$D$}^2 V\Big).
\end{equation}

\noindent (In our notation $\mbox{\boldmath$D$}$,
$\mbox{\boldmath$\bar D$}$, and $\mbox{\boldmath$D$}_\alpha$ are
the background covariant derivatives.) Certainly, the gauge fixing
procedure also requires introducing the Faddeev--Popov and
Nielsen--Kallosh ghosts. The higher covariant derivative
regularization can be also introduced without breaking the
background gauge invariance. This can be done by different ways.
For example, it is possible to add

\begin{eqnarray}\label{Regularized_Action1}
&& S_\Lambda = \frac{1}{2 e^2}\mbox{tr}\,\mbox{Re}\int
d^4x\,d^4\theta\, V\frac{(\mbox{\boldmath$D$}_\mu^2)^{n+1}}{
\Lambda^{2n}} V + \frac{1}{8} \int
d^4x\,d^4\theta\,\Bigg((\phi^*)^i
\Big[e^{\mbox{\boldmath${\scriptstyle \Omega}$}^+} e^{2V}
\frac{(\mbox{\boldmath$D$}_\alpha^2)^{m}}{\Lambda^{2m}}
e^{\mbox{\boldmath${\scriptstyle \Omega}$}}\Big]{}_i{}^j \phi_j
+\nonumber\\
&& + (\phi^*)^i \Big[e^{\mbox{\boldmath${\scriptstyle \Omega}$}^+}
\frac{(\mbox{\boldmath$D$}_\alpha^2)^{m}}{\Lambda^{2m}} e^{2V}
e^{\mbox{\boldmath${\scriptstyle \Omega}$}}\Big]{}_i{}^j
\phi_j\Bigg)
\end{eqnarray}

\noindent to the action, assuming that $n>m$. (It is important
that for a theory with a nontrivial cubic superpotential a term
with the higher covariant derivatives should be also introduced
for the matter superfields.)

As in the case of $N=1$ SQED, the higher covariant derivative term
does not remove divergences in the one-loop approximation, and in
order to regularize them the Pauli--Villars determinants should be
inserted into the generating functional. The Pauli--Villars fields
should be introduced for the matter superfields and all ghosts. (A
contribution of the gauge superfields in the one-loop
approximation vanishes.) Masses of the Pauli--Villars superfields
$\phi_I$ are proportional to the parameter $\Lambda$: $M^{ij}_I =
a_I^{ij}\Lambda$ and satisfy the relation $M_I^{ij} (M_I^*)_{jk} =
M_I^2 \delta_k^i$.

Using the background gauge invariance it is possible to choose
$\mbox{\boldmath${\Omega}$} = \mbox{\boldmath${\Omega}$}^+ = {\bf
V}$. Then the function $d^{-1}$ is defined by

\begin{equation}
\Gamma^{(2)}_V = - \frac{1}{8\pi} \mbox{tr}\int
\frac{d^4p}{(2\pi)^4}\,d^4\theta\,{\bf V}(-p)\,\partial^2\Pi_{1/2}
{\bf V}(p)\, d^{-1}(\alpha,\lambda,\mu/p).
\end{equation}

\noindent In order to find the $\beta$-function we again use
prescription (\ref{We_Calculate}) and find the derivative of
$d^{-1}$ with respect to $\ln\Lambda$ in the limit of the
vanishing external momentum. In the non-Abelian case the result is
given by

\begin{eqnarray}
&& \beta(\alpha,\lambda) = -\frac{3 \alpha^2}{2\pi} C_2 + \alpha^2
T(R) I_0 + \alpha^3 C_2^2 I_1 +
\frac{\alpha^3}{r} C(R)_i{}^j C(R)_j{}^i I_2  \nonumber\\
&& + \alpha^3 T(R) C_2 I_3 + \alpha^2 C(R)_i{}^j
\frac{\lambda_{jkl}^* \lambda^{ikl}}{4\pi r} I_4 +\ldots,
\end{eqnarray}

\noindent where

\begin{eqnarray}
&& \mbox{tr}\,(T^A T^B) \equiv T(R)\,\delta^{AB};\qquad
(T^A)_i{}^k
(T^A)_k{}^j \equiv C(R)_i{}^j;\nonumber\\
&& f^{ACD} f^{BCD} \equiv C_2 \delta^{AB};\qquad\quad r\equiv
\delta_{AA}.
\end{eqnarray}

\noindent The integrals defining the $\beta$-function are given by

\begin{equation}
I_i = I_i(0) -\sum\limits_I c_I I_i(M_I),\quad i=0,2,3,
\end{equation}

\noindent where for simplicity we do not write the ghost
constributions (they are also given by integrals of double total
derivatives) and

\begin{eqnarray}
&& I_0(M) = -\pi \int \frac{d^4q}{(2\pi)^4} \frac{d}{d\ln\Lambda}
\frac{\partial}{\partial q^\mu} \frac{\partial}{\partial q_\mu}
\Bigg\{\frac{1}{q^2}
\ln\Big(q^2(1+q^{2m}/\Lambda^{2m})^2+M^2\Big) \Bigg\};\\
&&\vphantom{1}\nonumber\\
&& I_1 = -12\pi^2 \int \frac{d^4q}{(2\pi)^4} \frac{d^4k}{(2\pi)^4}
\frac{d}{d\ln\Lambda} \frac{\partial}{\partial k^\mu}
\frac{\partial}{\partial k_\mu} \Bigg\{\frac{1}{k^2
(1+k^{2n}/\Lambda^{2n}) q^2 (1+q^{2n}/\Lambda^{2n})
(q+k)^2}\nonumber\\
&& \times \frac{1}{(1+(q+k)^{2n}/\Lambda^{2n})}  \Bigg\};\\
&&\vphantom{1}\nonumber\\
&& I_2(M) = 2\pi^2 \int \frac{d^4q}{(2\pi)^4}
\frac{d^4k}{(2\pi)^4} \frac{d}{d\ln\Lambda}
\frac{\partial}{\partial q^\mu} \frac{\partial}{\partial q_\mu}
\Bigg\{\frac{(2+(q+k)^{2m}/\Lambda^{2m}+ q^{2m}/\Lambda^{2m})^2
}{k^2 (1+k^{2n}/\Lambda^{2n})}
\nonumber\\
&& \times \frac{(1+ q^{2m}/\Lambda^{2m})(1+
(q+k)^{2m}/\Lambda^{2m})}{\Big(q^2(1+q^{2m}/\Lambda^{2m})^2
+M^2\Big)\Big((q+k)^2(1+ (q+k)^{2m}/\Lambda^{2m})^2 +M^2\Big)}
\Bigg\};\\
&&\vphantom{1}\nonumber\\
&& I_3(M) = 2\pi^2 \int \frac{d^4q}{(2\pi)^4}
\frac{d^4k}{(2\pi)^4} \frac{d}{d\ln\Lambda}
\frac{\partial}{\partial q^\mu} \frac{\partial}{\partial k_\mu}
\Bigg\{ \frac{(2+ k^{2m}/\Lambda^{2m} +
q^{2m}/\Lambda^{2m})^2}{(k+q)^2 (1+
(q+k)^{2n}/\Lambda^{2n})} \nonumber\\
&& \times
\frac{(1+k^{2m}/\Lambda^{2m})(1+q^{2m}/\Lambda^{2m})}{\Big(k^2(1+
k^{2m}/\Lambda^{2m})^2+M^2\Big)\Big(q^2(1+
q^{2m}/\Lambda^{2m})^2+M^2\Big)} \Bigg\};\\
&&\vphantom{1}\nonumber\\
&& I_4 = -8\pi^2 \int \frac{d^4q}{(2\pi)^4} \frac{d^4k}{(2\pi)^4}
\frac{d}{d\ln\Lambda} \frac{\partial}{\partial q^\mu}
\frac{\partial}{\partial q_\mu} \Bigg\{\frac{1}{k^2
(1+k^{2m}/\Lambda^{2m}) q^2 (1+q^{2m}/\Lambda^{2m}) (q+k)^2
}\nonumber\\
&& \times \frac{1}{(1+(q+k)^{2m}/\Lambda^{2m})} \Bigg\}.
\end{eqnarray}

\noindent Thus, we see that the $\beta$-function is given by
integrals of double total derivatives. In the considered
(two-loop) approximation they can be easily calculated
analytically using Eq. (\ref{Integral}):

\begin{equation}
\beta = - \frac{\alpha^2}{2\pi}\Big(3 C_2 - T(R)\Big) +
\frac{\alpha^3}{(2\pi)^2}\Big(-3 C_2^2 + T(R) C_2 + \frac{2}{r}
C(R)_i{}^j C(R)_j{}^i\Big) - \frac{\alpha^2 C(R)_i{}^j
\lambda_{jkl}^* \lambda^{ikl}}{8\pi^3 r} + \ldots
\end{equation}

\noindent This expression should be compared with the one-loop
anomalous dimension

\begin{equation}
\gamma_i{}^j(\alpha) = -\frac{\alpha C(R)_i{}^j}{\pi} +
\frac{\lambda_{ikl}^* \lambda^{jkl}}{4\pi^2} + \ldots
\end{equation}

\noindent Then we see that in the considered approximation the
$\beta$-function agrees with the exact NSVZ $\beta$-function

\begin{equation} \beta(\alpha) = - \frac{\alpha^2\Big[3 C_2 - T(R)
+ C(R)_i{}^j \gamma_j{}^i(\alpha)/r\Big]}{2\pi(1-
C_2\alpha/2\pi)}.
\end{equation}

\section{Conclusion}
\hspace{\parindent}

Although it is generally believed that the integrals appearing
with the higher covariant derivative regularization have too
complicated structure, we see that in the supersymmetric case some
of them can be calculated analytically. This makes possible
analytical multiloop calculations in supersymmetric theories with
this regularization. In principle, in the lowest loops it is not
very difficult to construct integrals corresponding to various
Green functions.

A very attractive feature of the higher covariant derivative
regularization is that all integrals defining the $\beta$-function
in the supersymmetric case seem to be integrals of double total
derivatives. As a consequnce, one of them can be calculated
analytically. (For $N=1$ SQED this was proved exactly in all
loops. For the general renormalizable $N=1$ SYM theory this was so
far verified only in the two-loop approximation.) In both
considered cases the factorization of integrands into total
derivatives allows to obtain the exact NSVZ $\beta$-function, and
for $N=1$ SQED for this purpose it is not necessary to redefine
the coupling constant. (In the non-Abelian case we cannot so far
make this conclusion, because the calculation was made only in the
two-loop approximation, where the $\beta$-function is
scheme-independent.)

\ack

This work was supported by Russian Foundation for Basic Research
grants No 11-01-00296-a and 11-02-08451-z. I would like to thank
the organizers of the ACAT conference for supporting my
participation. I am also very grateful to prof. A.L.Kataev for
valuable discussions.


\smallskip

\end{document}